# Complexation of a Thermoresponsive Brush-Type Polyelectrolyte with an Oppositely Charged Surfactant: Effect of Temperature and Surfactant Concentration.

Hernán A. Ritacco[(*)1,2], Macos D. Fernández Leyes[1], Zulma Quirolo[1], M. M. Soledad Lencina[1], Cecilia del Barrio[1], Rafael Márquez[2], Jaqueline Fernández[1], and Jhon Sánchez Morales[1].

[1] Instituto de Física del Sur (IFISUR-CONICET), Av. Alem 1253, Bahía Blanca (8000), Argentina

[2] Departamento de Física de la Universidad Nacional del Sur, Av. Alem 1253, Bahía Blanca (8000), Argentina.

(*) Corresponding author: hernan.ritacco@uns.edu.ar

**ABSTRACT.**

Responsive vectors for drug delivery could be formulated based on oppositely charged polyelectrolyte-surfactant complexes. As a model system, we synthetized a copolymer (PECop) made from alginate and Poly(N-isopropylacrylamide) PNIPAAm to obtain a brush-type polyelectrolyte with side chains capable of responding to temperature changes. We studied the aggregation process of PECop with a cationic surfactant, dodecyltrimethylammonium bromide (DTAB), as a function of surfactant concentration and temperature. We used surface tension, electrophoretic mobility and zeta ($\zeta$)-potential, potentiometry, dynamic and static light scattering, and atomic force microscopy (AFM) to characterize the copolymer/surfactant complexes.

We found that PECop/DTAB complexes are spherical and quite monodisperse in a certain surfactant concentration range, despite the broad polydispersity of the copolymer. The binding isotherms show a mixed behavior between a typical oppositely charged polyelectrolyte/surfactant mixture (sigmoidal) and hydrophobically modified polymers mixed with surfactants. We compared the binding isotherms of PECop with those of alginate and found that the number of DTAB molecules bound to the chains is 6 times higher in the former, when compared at the same total surfactant concentration (~1 mM).

The complexes respond to temperature changes, but only in certain small ranges of surfactant concentrations ($c_s$). Our results demonstrate that the surfactant molecules produce a progressive collapse of the polymer chains, which is maximum at $c_s$ = 2.8 mM, a concentration for which thermo-responsiveness is completely lost. For $c_s$ < 1 mM, PECop/DTAB aggregates decrease their size when temperature crosses the lower critical solution temperature (LCST) of the copolymer. On the contrary, for $c_s$ > 10 mM, the complexes increase their size when T > LCST. This inversion in the "sign" of the thermal response as the surfactant concentration increases indicates a change in the structure of aggregates and could open new possibilities in the design of nanocarriers for drug delivery systems using these kind of polymer/surfactant complexes.

# 1. INTRODUCTION

The formation of supramolecular aggregates of polymers and polyelectrolytes, including those of biological origin such as DNA or proteins, with other polymers and with surfactants has been studied for more than fifty years due to their potential in a variety of technological applications.[1]

Polyelectrolytes are polymers that split into macroions and small counterions when dissolved in water. Surfactants are, in general, small molecules whose chemical structure has two distinct parts: the polar head with affinity to polar solvents (water) and the hydrophobic tail with affinity to nonpolar fluids. These molecules have the property of adsorbing spontaneously at the interface separating two immiscible fluids, one polar and one nonpolar, such as the air-water interface. At a certain concentration, surfactants self-aggregate in bulk to form micelles [2], the concentration at which this happens is called critical micelle concentration (*cmc*). Polyelectrolytes and surfactants are used in a broad number of technological applications both on their own and mixed. The behavior of polyelectrolyte/surfactant mixtures is so reach[3] that they are considered as the platform for a large number of different new technological applications. Some of these include paints, wastewater treatment, the personal care and oil industries, as gene carriers in gene therapy, and encapsulation in drug delivery systems, among others.[4–9] In this sense, pH-responsive and thermoresponsive systems are of particular interest in nanomedicine[10] because of the environmental conditions found in the human body.

We are concerned here with the association between polyelectrolytes and oppositely charged surfactants.[3–5,11] In this case, the association between species is driven by both hydrophobic and electrostatic attraction. The features of the complexes formed and the phase behavior of the solutions are the result of an intricate balance between attractive and repulsive interactions between polyelectrolytes and surfactants. These interactions depend on the chemical nature of surfactants and polyelectrolytes, such as charge density, hydrophobicity of chains, molecular weight, degree of branching, etc., as well as on the concentration of polyelectrolytes and surfactants, but also on the physical conditions, such as ionic strength, pH, or temperature. This dependence on physical parameters opens the door to the formulation of systems capable of responding to an external trigger, for example, changes in pH or temperature.

For the sake of context, an oversimplified picture of the behavior of polyelectrolyte/surfactant mixtures is given below.[11] First, when an oppositely charged surfactant is added to a polyelectrolyte solution, it progressively replaces the polyelectrolyte counterions in the vicinity of the macromolecular main chain. This process is mainly driven by entropy gain and does not lead to observable changes in the bulk properties of the system, as might be observed with commonly used techniques such as conductivity or light scattering; however, they can be detected by less common techniques such as electric birefringence.[12–16] This situation changes when a certain surfactant concentration, the critical aggregation concentration (*cac*), is reached. At this concentration, surfactant molecules cooperatively bind onto the macromolecule chain. In general, *cac* occurs at concentrations one to three orders of magnitude lower than the *cmc* of the surfactant and can be determined by calorimetry, conductivity or surface tension[1] measurements. In this last technique, *cac* is ascribed to the beginning of the first plateau in the surface tension isotherms;[17] this concentration is also known as T1. As the surfactant concentration continues to increase, surface tension remains almost constant (plateau) until a concentration T2 is reached. Generally, T2 concentration is ascribed to the saturation of the binding sites on the polyelectrolyte chain. At this point, surfactant/polyelectrolyte complexes become hydrophobic and phase separation may occur. At higher surfactant concentrations, generally above the *cmc* of the pure surfactant, redissolution of these precipitates may happen. This last concentration is commonly referred to as T3. The features of the polymer/surfactant complexes, such as size and form, have been studied with a number of techniques including dynamic light scattering (DLS) and static light scattering (SLS), X-ray spectroscopy, small-angle X-ray scattering (SAXS) and small-angle neutron scattering (SANS), among others.[11,18–22] It was found that a large number of factors influence the size and form of polyelectrolyte/surfactant complexes as well as the characteristics of phase diagrams.[23,24] To make things even more complicated, polyelectrolyte-surfactant complexes often remain trapped in nonequilibrium metastable states whose characteristics depend on the history of the systems, for instance, mixing protocols or the time elapsed since preparation.[25–29]

In the present article, we study a mixture of a copolymer, a brush-type polyelectrolyte, hereinafter referred to as PECop, and a cationic surfactant, dodecyltrimethylammonium bromide (DTAB). The copolymer was formed from alginate, which is a natural polyelectrolyte widely used as a gelling agent in food[30], pharmaceutical applications[31,32] and smart materials for encapsulation[33], and Poly(N-isopropylacrylamide) (PNIPAAm), which is incorporated as side chains. PECop was synthetized under the hypothesis that the brush side chains could increase the binding sites for the small molecules, and thus the cargo capabilities for drug delivery, while the alginate backbone could respond to pH and PNiPAAm side chains to temperature. Therefore, PECop sets up a potentially double response system for the formulation of smart materials. In a previous work, we studied this system at interfaces within the framework of thermoresponsive foams; here, we focus on bulk properties and responsiveness to temperature. Our aim is to study the polyelectrolyte-surfactant assembling process and the structure of aggregates in bulk, their dependence on surfactant concentration and temperature, as well as to evaluate how the responsiveness to temperature depends on those variables. This last feature is fundamental to successfully formulate useful technological systems for drug delivery based on mixtures of polyelectrolytes with oppositely charged small molecules.

## 2. MATERIALS AND METHODS

### 2.1 Materials

Dodecyltrimethylammonium bromide (DTAB), a cationic surfactant, and Sodium dodecyl sulfate (SDS), an anionic surfactant, were purchased from Sigma-Aldrich (>99%) and used as received. Sodium alginate is the sodium salt of alginic acid, a linear polysaccharide obtained from brown algae. A low viscosity sodium alginate was purchased from Sigma-Aldrich (Argentina). PNIPAAm is a synthetic polymer that presents a lower critical solution temperature (LCST). PNIPAAm undergoes a conformational transition from random coils to globules when temperature surpasses the LCST (~32 °C for high molar mass). LCST is a function of molar mass and polymer concentration, among other parameters.[34–38]

The methodology we followed for the synthesis of the alginate-g-PNIPAAm graft copolymer (PECop) is well known.[39] It consists of a coupling reaction between the carboxyl groups of sodium alginate and the terminal amine groups of PNIPAAm-$NH_2$ chains, using 1-ethyl-3-(3´-(dimethylamino) propyl) carbodiimide hydrochloride (EDC) as the coupling agent. We synthesized a brush-type anionic polyelectrolyte with Mn = 4200 g/mol PNIPAAm side chains. The mean molecular weight of the copolymer was determined by SLS, giving a value of Mw = 89.5 kDa (PD ~ 2). The total number of monomers per copolymer chain is ~514, being the molar fraction of PNIPAAm repeating units of about 28%. Thus, the number of charges per copolymer molecule was found to be about 370 (~ 1.65 mM for a polymer concentration of 400 ppm).

We used ultrapure water (Milli-Q water purification system) for the preparation of all polyelectrolyte and surfactant solutions. A single, fixed polymer concentration ($c_p$) of 400 mg L$^{-1}$ was used in the preparation of almost all samples, except for electrophoretic mobility experiments in which, in order to obtain binding isotherms (see following sections), we also used solutions at a polymer concentration of $c_p$ = 100 mg L$^{-1}$.

### 2.2 Sample Preparation Protocols and Measurements.

Two different sample preparation protocols were used. For surface tension measurements, a concentration process was employed. First, the surface tension of a DTAB-free aqueous solution of PECop at $c_p$ = 400 mg L$^{-1}$ was measured. Subsequently, we added proper amounts of the copolymer and DTAB solutions until the targeted concentration was reached; the surface tension was then measured after an equilibration period of 60 min. This process was repeated until the whole range of DTAB concentration was covered.

For dynamic (DLS) and static (SLS) light scattering, mobility and zeta ($\zeta$)-potential measurements, all samples were obtained by adding equal volumes of DTAB solution with twice the desired final concentration to 800 mg L$^{-1}$ of the PECop solution so that the final polymer concentration was always 400 mg L$^{-1}$ (or 100 mg L$^{-1}$). Solutions were left to reach equilibrium for 24h prior to measurement. Some of these bulk experiments, DLS and $\zeta$-potential, were repeated with samples prepared following the first preparation protocol; there were no significant differences in the corresponding results.

### 2.3 Methods

### 2.3.1 Surface Tension

Surface tension measurements were carried out using the sensor of a Langmuir balance (KSV NIMA medium) and a Wilhelmy paper plate. Experiments at room temperature (~22-25 °C) were performed using a Teflon trough (10 ml volume). A jacketed vessel was employed for temperature-dependent measurements.

Pure water surface tension measurements were used to verify optimal paper probe quality prior to each experimental iteration. After pouring the solutions into the corresponding vessel, surface tension was continuously measured until a stable value was reached. The reproducibility was ± 0.2 mN m$^{-1}$.

Temperature-dependent experiments were performed in the range of 20 °C to 55 °C, with measurements every 5 °C. An approximate heating rate of 1 °C/min was used between steps. Once the required temperature was reached, samples were left to equilibrate for at least 30 min before surface tension determination. Temperature was controlled using an external circulating water bath (Lauda Alpha R8) and monitored by a thermocouple.

### 2.3.2 Dynamic (DLS) and Static (SLS) Light Scattering

The hydrodynamic diameter of the complexes (that is, the diameter of a compact, homogeneous sphere that has the same translational diffusion coefficient as the aggregate) were measured as a function of temperature and DTAB concentration by DLS. The majority of DLS measurements were performed in a Malvern Zetasizer Nano ZSP (light source 10 mW He-Ne laser, wavelength 633 nm) from Malvern Instruments, at a scattering angle (θ) of 13°. The intensity autocorrelation functions were analyzed using Cumulants [40] or CONTIN[41] analysis, from which we obtained the apparent translational diffusion coefficients, $D_s$. Examples of the autocorrelation functions can be seen in Figure SI-1 in the supporting information document (SI).

Once $D_s$ was obtained, the hydrodynamic diameter, $D_H$, was determined from the Stokes-Einstein equation,

$$D_s = \frac{k_B T}{3\pi \eta D_H} \quad (1)$$

being $k_B$ the Boltzmann constant, T the temperature, and η the solvent viscosity. The temperature was controlled (± 0.1 °C) using the own system of the device.

Static light scattering, that is, the intensity of light scattered by the samples at each scattering angle, was measured in a Malvern 4700 Multiangle apparatus, as a function of the scattering vector, q, $q = \frac{4\pi n \sin(\frac{\theta}{2})}{\lambda}$, at angles between 20° and 150°, by steps of 1°. The measured intensities were converted to the absolute scale using pure toluene as the scattering standard.

The corresponding dependence of the scattered intensity on q, I(q) (and form factors, P(q) ~ I(q)), was analyzed using the Guinier-Porod empirical law, [42–44]

$$I(q) \sim \frac{1}{q^s} \exp\left[-\frac{q^2 R_g^2}{3-s}\right] \quad for\ q \leq q_l = \frac{1}{R_g}\sqrt{\frac{(m-s)(3-s)}{2}}$$

$$I(q) \sim \frac{1}{q^m} \quad for\ q \geq q_l \quad (2)$$

where m is the Porod exponent, $R_g$ is the radius of gyration, and s is a dimensional variable (for 3D globular objects, such as spheres, s = 0; for 2D symmetry, such as rods, s = 1, and for 1D objects, such as lamellae or platelets, s = 2).

### 2.3.3 Electrophoretic Mobility and ζ-Potential

The electrophoretic mobility and ζ-potential[45] of polyelectrolyte/surfactants aggregates were measured with a Malvern Zetasizer Nano ZSP (light source 10 mW He-Ne laser, wavelength 633 nm) from Malvern Instruments. This instrument uses the laser Doppler velocimetry method with phase analysis light scattering (PALS) to obtain the electrophoretic velocity, v, of colloidal particles and from it the mobilities, $u = v/E$, where E is the applied electric field. Once $u$ is measured, the ζ-potential is calculated using the Henry equation and Smoluchowsky approximation, $\zeta = \eta\,(u/\epsilon)$, where η is the solvent viscosity and ε is the solvent permittivity, respectively.

Each mobility value obtained is an average of several measurements, according to Malvern´s proprietary "quality factor" statistical criterion.

Disposable capillary cells were used. Samples were allowed to reach their equilibrium temperature for 60 min prior to experiments. Values were taken in triplicate with a delay of 120 s in between.

### 2.3.4 AFM

AFM (Bruker Innova) measurements were performed under ambient conditions in tapping mode using RTESP-CP tips (Veeco, spring constant = 20-80 N/m, as reported by manufacturer). Samples were prepared by casting drops of solutions containing copolymer/surfactant mixtures onto a smooth glass surface and then evaporating the water in a vacuum chamber. Images with a scan range of 3 μm at a scan rate of 1 Hz were taken and processed using the Gwyddion software.

### 2.3.5 Viscosity Measurements

The viscosities of aqueous solutions of selected complexes with DTAB concentration of $c_s$ = 0; 0.3; 1.62; 2.82; and 15 mM were determined at 25 °C and 45 °C using an Ubbelohde viscometer. The values reported are the average of 10 measurements.

### 2.3.6. Potentiometry using surfactant-selective electrodes

The equilibrium concentration of DTAB ions, DTA+ in the polyelectrolyte solutions was determined potentiometrically with a DTA+ selective plastic membrane electrode. We constructed the surfactant ion selective electrode as described by Hayakawa et al.[46]. First, we precipitated the insoluble DTA·DS complex by mixing equal amounts (25 mL) of DTAB and SDS 0.02 M. The solid formed was filtered using a Whatman (Double ring Qualitative filter Paper) paper filter, washed with water and left to dry. Then, we made the membrane with the following composition: 25% of PVC (Poly(vinyl chloride)high molecular weight Aldrich, 0.5011g), 74% of plasticizer (Dioctyl phthalate ≥99.5% Aldrich, 1.4828g) and 1% of the insoluble salt (DTA·DS, 0.0221g) by dissolving all the components in tetrahydrofuran (THF) and left to evaporate in a 10 cm Petri dish. The obtained membrane was glued to a PVC tube. To perform measurements, we filled the tube with DTAB aqueous solution (1 mM) and completed the system with two identical Ag/AgCl reference electrodes from Van London, one inside the tube and the other in contact with the solutions to be measured. We measured the electric potentials with a Jenco 6177 millivoltmeter (± 0.1 mV precision, ± 1 mV accuracy). The response remained linear to concentration as low as 0.01 mM.

## 3. RESULTS

### 3.1 Phase Behavior

The phase behavior of the PECop/DTAB mixed solutions was observed as a function of temperature and surfactant concentration by simple naked eye observation (see Figure SI-2). At T = 20 °C and for all surfactant concentrations from 0 to 30 mM, the suspensions are stable, and no phase separation are observed after 48 h. As the temperature increases from 20 °C to 55 °C, phase separation is observed for surfactant concentrations between 8 mM and 15 mM. Below and above this concentration range, the systems are stable (no precipitate) at all temperatures.

For comparison, we performed the same observations for alginate/DTAB mixtures. The alginate used is from the same batch used for the synthesis of PECop. The solutions were stable for at least 48 hs, without precipitation, at surfactant concentrations ($c_s$) below 8.6 mM and above 15 mM, at 25 °C. For 8.6 < $c_s$ < 15 mM we observed precipitation. Note that this behaviour is different from PECop/DTAB complexes that do not precipitate in the whole surfactant concentration explored when T<LCST. The alginate/DTAB precipitates do not redissolve at T= 45 °C (see Figure SI-3).

### 3.2. Equilibrium Surface Tension Isotherms

Surface tension (γ) measurements were performed in several aqueous solutions with increasing DTAB concentration ($c_s$) and a fixed PECop concentration, $c_p$= 400 mg L$^{-1}$. Measurements were performed at two temperatures, 25 °C and 45 °C. Results are shown in Figure 1. Note the significant drop in surface tension caused only by the copolymer (PECop), displaying a clear surface activity.

The surface tension of DTAB-free PECop solutions were 45.8 mN.m$^{-1}$ and 39 mN.m$^{-1}$, at 25 °C and 45 °C, respectively.

Regarding the effect of DTAB on surface tension, Figure 1 shows the presence of two plateaus. For the measurements at T = 25 °C, the first plateau begins at a surfactant concentration of approximately $c_s \sim 0.7$ mM (T1 in the figure) and ends at approximately $c_s \sim 7$ mM (T2 in the figure). Then, as $c_s$ increases, the surface tension decreases until the second plateau begins at T3, which occurs at $c_s \sim 17$ mM, close and above the *cmc* of pure DTAB solutions (~15 mM). Thereafter, the surface tension remains constant up to the highest surfactant concentration used, $c_s \sim 80$ mM. A similar behavior is observed for T = 45°C, for which all T1, T2, and T3 shift slightly to lower concentrations. Concentrations are as follows: T1 ~ 0.5 mM, T2 ~ 6 mM, and T3 ~ 15 mM. No peaks in surface tension isotherms are observed after allowing the system to equilibrate for 48 hours.[47–50]

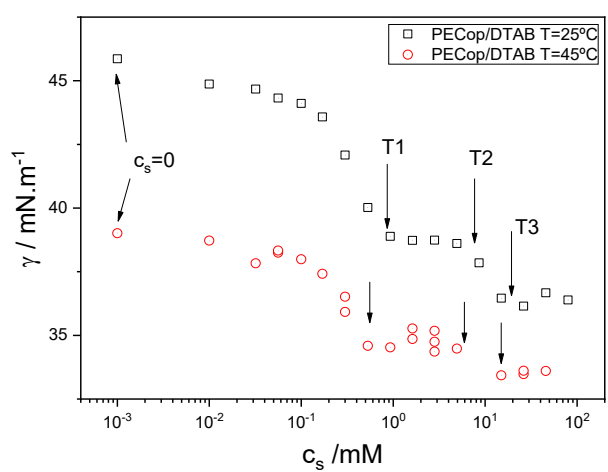

**Figure 1.** Surface tension of PECop/surfactant mixtures as a function of DTAB concentration at 25 °C and 45 °C.

### 3.3 Viscosity

Figure 2 shows relative viscosity (with respect to water at the same temperature) as a function of surfactant concentration. Note that viscosity is higher at T = 25 °C than at T = 45 °C, as expected, if $c_s < 2.8$ mM. At this surfactant concentration, a transition occurs. At $c_s \geq 2.8$ mM, the viscosity of the solutions becomes the same as that of water at the same temperature.

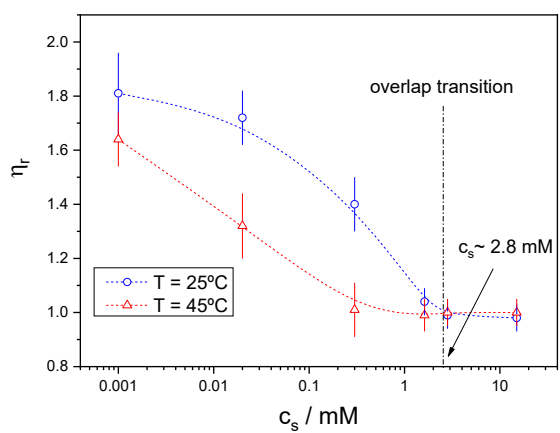

**Figure 2**. Relative viscosity as a function of surfactant concentration at two different temperatures. T = 25 °C (blue circles); T = 45 °C (red triangles).

### 3.4 DLS and SLS: Size and Form of Aggregates

To obtain information on the size of the aggregates, we performed DLS experiments. We measured the hydrodynamic diameter, $D_H$, at the smallest scattering angle available in our light scattering apparatus, $\theta = 13°$. We first measured $D_H$ as a function of temperature for the polyelectrolyte alone, as shown in Figure 3. A sharp transition temperature, LCST, was found at $38 \pm 1$ °C, with $D_H$ ranging from about 1300 nm, below LCST, to 450 nm, above it. In these samples, the correlation functions were well fitted with monoexponentials (see Figure-SI-1), at least in the time range explored. For future comparison purposes, we show in Figure 4a the hydrodynamic diameters, $D_H$ as a function of DTAB concentration at T= 25 °C for the complexes of alginate, the same batch used in the synthesis of the copolymer PECop, with DTAB. $D_H$ values change from about 700 nm at very low surfactant concentrations to 25 nm at $c_s$= 1.6 mM. At $c_s$> 1.6, $D_H$ continuously increases with surfactant concentration. We added in Figure 4a a dashed zone indicating the region where precipitation of alginate/DTAB complexes are observed. The precipitation zone begins at $c_s$= 8.6 mM and ends at $c_s$ = 15 mM. Figure 4b presents the hydrodynamic diameters, $D_H$ as a function of DTAB concentration, for two temperatures, above and below the LCST temperature for the PECop/DTAB complexes.

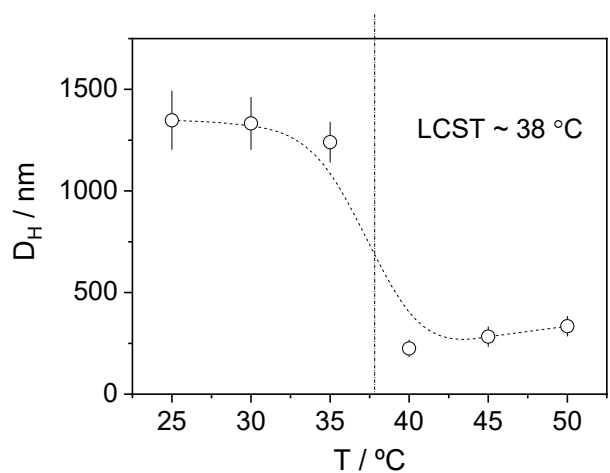

**Figure 3.** Hydrodynamic diameters, $D_H$, for PECop 400 ppm without surfactant. The scattering angle was set to 13°.

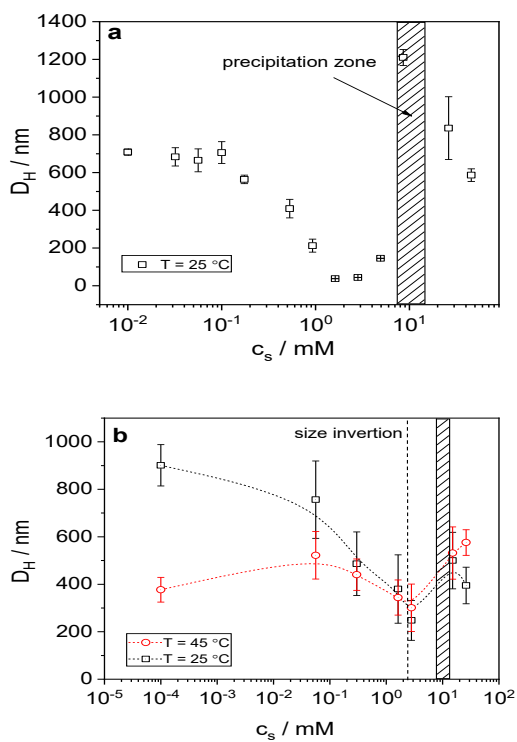

**Figure 4. (a)** $D_H$ for Alginate/surfactant complexes in aqueous solution, $c_p$ = 400 mg L$^{-1}$, as a function of DTAB concentration at 25 °C; **(b)** $D_H$ for PECop/surfactant complexes in aqueous solution, $c_p$ = 400 mg L$^{-1}$, as a function of DTAB concentration at 25 °C (black squares) and at 45°C (red circles)

Note that $D_H$ decreases by a factor of about 2 as temperature becomes higher than LSCT for all mixtures with $c_s < 0.1$ mM. For $0.3 < c_s < 2$ mM, the hydrodynamic diameters seem to be quite close at both temperatures, given the errors bars, being the aggregates slightly larger for lower temperature. At $c_s = 2.8$ mM, the sizes of the aggregates are equal at both temperatures. For $c_s > 10$ mM, the behavior is clearly reversed, the response changes "sign", and $D_H$ increases as the temperature rises from 25 °C to 45 °C and surfactant concentration increases. This behaviour is also observed for alginate/DTAB complexes (Figure 4a). The polydispersity index (PI) obtained from the cumulant analysis of the intensity autocorrelation functions is between 0.05 and 0.5 for all samples at concentrations $c_s < 3$ mM, both at T = 25 °C and 45 °C. These results indicate the existence of rather monodisperse aggregates, particularly when T > LCST.

We also performed SLS measurements to obtain information on the form of these aggregates. The intensity of scattered light as a function of the scattering angle (SLS) is shown in Figures 5 and 6. The concentrations studied were those corresponding to $c_s = 0$ (pure PECop) at T= 45°C, $c_s = 1.6$, and $c_s = 2.8$ mM at T=25°C. We measured only these concentrations due to a limited amount of PECop and because they are in the maximum collapsed state, which is more suited for the determination of the shape of the aggregates by SLS. Figure 5 shows the intensity of scattered light as a function of wave vector q (form factor), for a mixture with $c_p = 400$ mg L$^{-1}$ and $c_s = 1.6$ Mm, at T = 25 °C. The line in the figure corresponds to the fitting curve with the Guinier-Porod empirical model, eq. 2, which gave $R_g = 184$ nm, s=0.27 and m=3, these values are compatible with spherical particles. The ρ-ratio, $\rho = R_g/R_H = 184/191 \sim 0.96$, is again compatible with spherical aggregates, which is confirmed by AFM images (inset Figure 5).

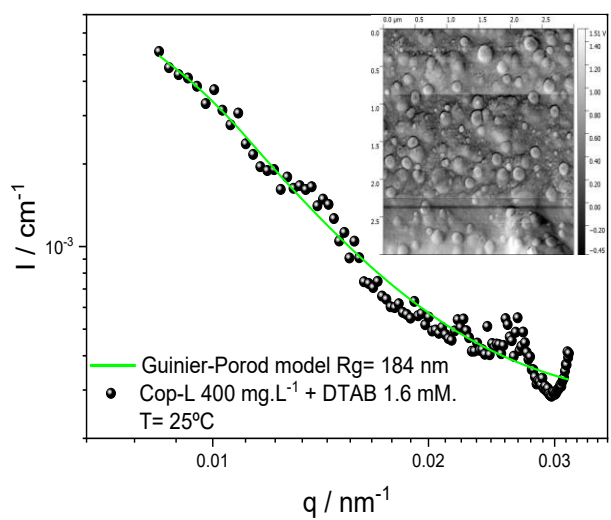

**Figure 5.** Experimental form factor (circles) for the systems PECop 400 mg.L$^{-1}$ + DTAB 1.6 mM at 25 °C. The line corresponds to the fitting with Guinier-Porod model (see text). The inset is an AFM image obtained for a PeCop 400 ppm/DTAB 1.6 mM mixture deposited on a Si-wafer and dried.

Figure 6a shows the form factor for a PECop solution with $c_p = 400$ mg L$^{-1}$ without surfactant at T = 45 °C and for a mixture with $c_p = 400$ mg L$^{-1}$ and $c_s = 2.8$ Mm at T = 25 °C (Figure 7b). In the figures, we include fittings with the Guinier-Porod empirical law (eq.2). From the fittings, $R_g = 170$ nm for the former solution. Because multiple scattering is present and the Rayleigh-Debye-Gans limits are not fulfilled in this case ($qR_g < 1$), the results should be taken with caution; however, the ρ-ratio, $\rho = R_g/R_H = 170/187 \sim 0.9$, is consistent with spheroidal aggregates. The same can be said for $c_s = 2.8$ mM, from the fittings with eq. (2), $R_g$ = 167 nm, $\rho = R_g/R_H$ =167/123 ~1.4 (ellipsoid). The values found for the Porod exponent, m, and for the dimensional parameter, s, are compatible with globular aggregates with fractal surfaces (m ~ 2.5, s ~ 0.5) for both solutions.

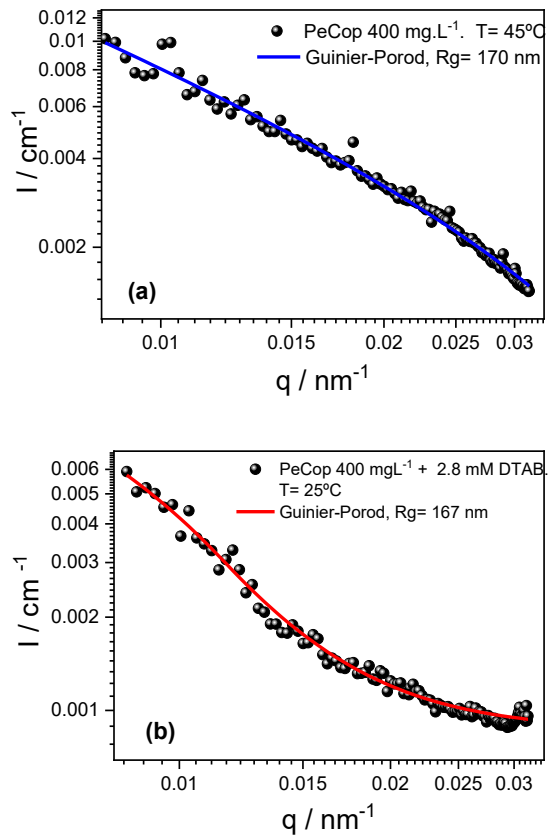

**Figure 6.** Experimental form factor (circles) for the systems (a) PeCop 400 mg L$^{-1}$ at 45 °C; (b) PeCop 400 mg L$^{-1}$ + 2.8 mM DTAB. The lines correspond to the fitting with the Guinier-Porod empirical model.

### 3.5 Electrophoretic Mobility Measurements

Figure 7a shows results of electrophoretic mobility for the mixed PECop/surfactant system for two different copolymer concentrations at 25 °C. Figure 7b shows the ζ-potential as a function of surfactant concentration for two temperatures, above and below LCST. The curve for T = 45 °C is qualitatively similar to that at 25 °C, except for the small region where a precipitate appears, indicated by a bar in the figure.

Note that the mobility and ζ-potential become zero at a total surfactant concentration of about 15 mM ($c_p$ = 400 mg L$^{-1}$), which coincides with the surfactant *cmc*.[51] It has been shown that, under certain conditions, the amount of surfactant molecules bound to the polyelectrolyte can be estimated from electrophoretic mobility measurements[52] for two different polymer concentrations; later on, we will use the results in Figure 7a with that purpose (see discussion below).

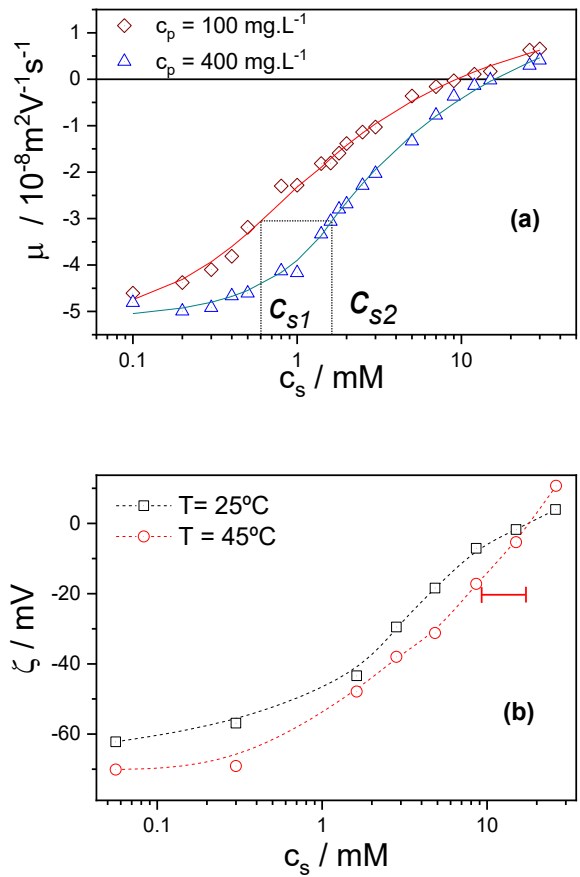

**Figure 7.** (a) Electrophoretic mobility of PeCop/DTAB complexes versus DTAB concentration. Solid lines correspond to the ad hoc fitted functions used to interpolate $u$ values. Dotted lines indicate schematically how DTAB concentration belonging to equal values of $u$ is determined (see discussion). (b) $\zeta$-potential from mobility measurements as a function of DTAB concentration at two temperatures, below and above LCST.

## 4. DISCUSSION.

### 4.1 Phase Behavior and Surface Tension

The surface activity shown in Figure 1 by the surfactant-free solutions, $c_s = 0$, is mainly attributed to the presence of PNIPAAm side chains in the copolymer, since alginate aqueous solutions show no significant surface activity at similar concentrations.[53] In this sense, Zhang et al.[54] reported a sizable effect of PNIPAAm on the surface tension of aqueous solutions, even for concentrations as low as 5 mg L$^{-1}$.

From the surface tension isotherms (Figure 1), we identified three characteristic surfactant concentrations that we called T1, T2, and T3. T1, which corresponds to the beginning of the first plateau, is generally associated with *cac* and corresponds to the onset of the binding of DTAB to PECop in bulk. Upon further increasing the amount of surfactant, the plateau ends at a concentration T2. At this point, it is generally assumed that all polymer binding sites are occupied by surfactant molecules and any excess causes a decrease in surface tension until the surfactant *cmc* is reached. Note that T2 (~ 7 mM) is below the concentration at which the electrophoretic mobility approaches zero (~ 15 mM; see Figure 7a), which is close to the surfactant *cmc* and just below T3 in the surface tension isotherm. Above T3, any DTAB addition would lead to the formation of free micelles, with no effect on surface tension.[55] Besides the overall decrease in surface tension previously mentioned, the increase in temperature seems to cause a slight shift of T1, T2, and T3 towards lower concentrations, probably due to the collapsed state of PNIPAAm side chains that lead to an increased hydrophobicity interaction between polymer and surfactant. Furthermore, in contrast to the behavior observed at 25 °C, at 45 °C the polymer precipitates in a concentration region between 8 mM and 15 mM, that is, between T2 and T3, which is also attributed to the increased hydrophobicity of the aggregates at higher temperature. Note that this surfactant concentration range is the same found for precipitation in alginate/DTAB mixtures. At concentrations above T3, the precipitates redissolve, leading to stable dispersions. This last concentration coincides with the surfactant *cmc* and with the surfactant concentration region where a size increase is observed as temperature increases above the transition temperature (see DLS data); thus, we interpret this as indication of a change in the structure of the aggregates. We attribute this to the formation of surfactant micelles attached to the polymer chain (note that this also happens for alginate/DTAB systems at the same surfactant concentration, see Figure 4a).

## 4.2 DLS and SLS

The results of DLS contrast, and are in many aspects opposite, to those of the PNIPAAm /SDS system.[56] The addition of sodium dodecyl sulfate (SDS) to PNIPAAm solutions produces a monotonous increase in hydrodynamic diameter both at temperatures below and above LCST, which is opposite to the behavior of PECop/DTAB complexes that undergo a progressive reduction in $D_H$ as DTAB concentration increases. Authors ascribe the behavior of PNIPAAm/SDS complexes to the electrostatic repulsion between the SDS molecules bound to the polymer chain that prevents the coil-globule transition. In fact, for this system, at a surfactant concentration above 350 mg L$^{-1}$ and for T > LCST, the addition of surfactant produces a globule-to-coil transition. The addition of DTAB to PECop produces a coil-to-globule transition when $c_s$ < 2.6 mM, the transition occurs because surfactant molecules bind either to the alginate backbone or to PANIPAAm side chains, but in view of ζ-potential results (Figure 7b), they do so retaining some of their counterions. Furthermore, for PNIPAAm/SDS the addition of SDS to PNIPAAm does not change the hydrodynamics diameter, both at temperatures above and below LCST, until a critical SDS concentration is reached (~ 350 mg L$^{-1}$), indicating a highly cooperative binding process. On the contrary, for PECop/DTAB, $D_H$ diminishes continuously and progressively (for $c_s$ < 2,8 mM), indicating a noncooperative binding process. This behaviour is also observed for alginate/DTAB complexes (Figure 5a) for 0.2 <$c_s$< 1.6 mM. Finally, the change of the "sign" of the response with temperature at $c_s$ > 10 mM when crossing the LCST observed for PECop/DTAB complexes is not observed for PNIPAam/SDS aggregates.

The results of DLS for PECop/DTAB and alginate/DTAB complexes shown in Figure 4 are similar in certain aspects to those found for DTAB/CarboxyMC (sodium corboxymethylcellulose).[57] The addition of an oppositely charged surfactant to a flexible polyelectrolyte produces, at certain concentrations, the polymer collapse. This results in aggregates that are spherical and monodisperse, as evidenced by DLS and SLS results (Figures 4 and 5). The monodispersity of the aggregates is quite surprising, considering that the size distribution of the polyelectrolyte chain is rather broad. [57]

The hydrodynamic diameters plotted in Figure 4b clearly show that the addition of DTAB produces a collapse of the polymer chain in a way similar to that produced by an increase of temperature above LCST for DTAB-free PECop solutions. By comparing with the DLS results for alginate/DTAB, which has a similar behaviour (Figure 4a), it seems that DTAB molecules bind mainly to the alginate backbone, probably close to the charged groups. However, ζ-potential results indicate that, if they do, some of the DTAB molecules retain their counterions after binding (ζ-potential ~ 0 only at $c_s$ = 15 mM). This collapse of the polymer chain as DTAB concentration increases is also observed by viscosity measurements (Figure 2); at 25 °C, the viscosity relative to water, $\eta_r = \frac{\eta_{solution}}{\eta_{water}}$, change from about 2 to 1 when surfactant concentration goes from 0 mM to 2.8 mM. At all surfactant concentrations over 2.8 mM, the viscosity of the solutions is that of pure water (Figure 2).

The effect of increasing aggregate size after collapse as DTAB concentration increases was also observed in DTAB/carboxyMC, [57] and in alginate/DTAB (Figure 4a). This suggests a change in the structure of the aggregates as $c_s$ increases above T2 (Figure 4) and probably a charge inversion of the colloid, a conclusion supported by ζ-potential results. In this respect, we tried to study the structure of the aggregates by small angle X-ray scattering (SAXS); unfortunately, the contrast condition was so poor that we did not obtain any usable results. This indicates that the aggregates are probably full of water.

### 4.3 Electrophoretic Mobility, ζ-Potential, and Binding Isotherms

In Figure 8 we show the binding isotherm for alginate/DTAB mixtures measured by potentiometry using surfactant-selective electrodes, which will be used for comparison purposes in the discussion that follows. The alginate used for these experiments is from the same batch used to synthesize PECop.

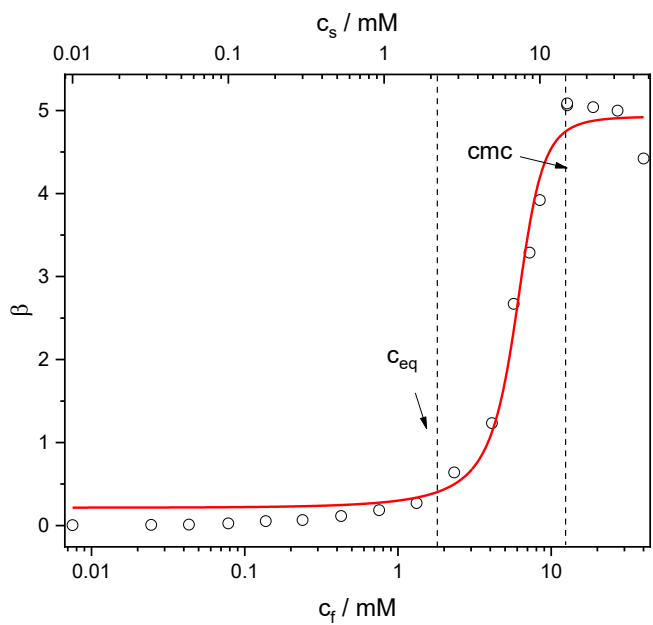

**Figure 8.** Binding isotherm measured using a surfactant-selective electrode for alginate/DTAB mixtures (open circles). The line is a fitting using Equation (6), see text.

Unfortunately, the available amount of PECop was insufficient to perform the measurement of binding isotherms by potentiometry using surfactant-selective electrodes because the volumes required for this technique are too large (~40 mL). However, according to Mezei et al.,[52] the binding isotherms of ionic surfactants on oppositely charged polymers can be estimated from electrophoretic mobility data of surfactant/polymer complexes. The relative number of surfactants bound to the polymer, B, can be expressed as

$$B = \frac{c_s - c_f}{c_p} \quad (3)$$

where $c_s$ is the total surfactant concentration, $c_f$ is the equilibrium free surfactant concentration, and $c_p$ is the polymer concentration. Equation (3) is valid if $c_f$ is lower than *cmc* and if either the ionic strength of the solution is high, or the charge density and the polymer concentration are not too high.[52,58]

To calculate B it is assumed that it depends only on $c_f$, and that the electrophoretic mobility $u$ is a function of B and is independent of the polymer concentration. For two polymer concentrations, $c_{p1}$ and $c_{p2}$, the same $B(c_f)$ and thus the same electrophoretic mobility $u$, will be achieved at two different surfactant concentrations, $c_{s1}$ and $c_{s2}$:

$$B(c_f) = \frac{c_{s1} - c_f}{c_{p1}} = \frac{c_{s2} - c_f}{c_{p2}} \tag{4}$$

$$u(c_{p1}, c_{s1}, c_f) = u(c_{p2}, c_{s2}, c_f) \tag{5}$$

The determination of $c_f$ is possible by means of eq. 4, using interpolated values of $c_s$ corresponding to equal mobilities, according to eq. 5.

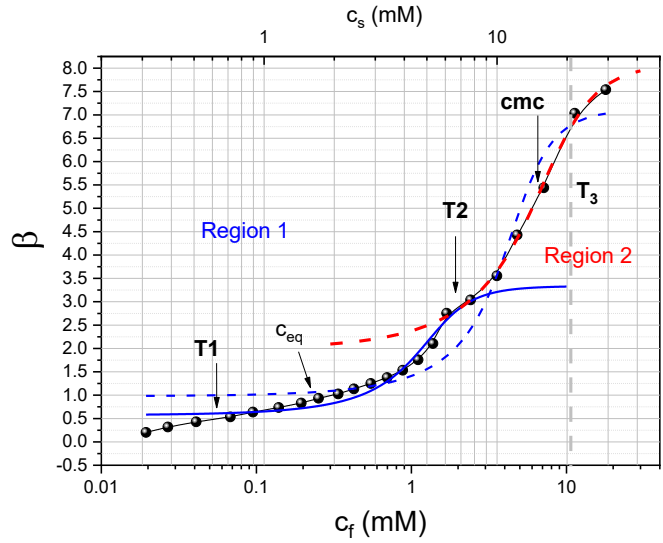

**Figure 9.** Binding isotherm of DTAB; β is the number of DTAB molecules bound divided by the number of charged groups in the polyelectrolyte chain, and $c_s$ and $c_f$ are the total and free (not bound) DTAB concentrations, respectively. Concentrations T1 and T2 obtained from surface tension isotherms are marked with arrows. The lines are fittings with eq. 6, for Region 1 (blue) and 2 (red) as well as for the whole isotherm (dashed line) (see text).

Figure 7a shows the mobilities measured at two different polymer concentrations, and Figure 9 shows the corresponding binding isotherm calculated from them using the procedure just outlined. From the mobilities, we obtained the free surfactant concentration but, in order to calculate the degree of binding, β, defined as β = (bound surfactant/binding sites on the polyelectrolyte), it is necessary to know the number of binding sites on the polymer chain. In general, for polyelectrolyte/oppositely charged surfactant mixtures, the number of charged groups on the polymer chain is assumed to be the number of binding sites. However, this assumption is not necessarily correct, especially if other types of interactions, besides electrostatic ones, are involved. For example, in the case of hydrophobically modified polymers (HMP), which are water-soluble polymers with low proportions of hydrophobic side chains, is necessary to consider two stoichiometries when describing the binding isotherms of their complexes with surfactants: the hydrophobic and charge stoichiometries.[59] PECop is not exactly an HMP, but being a brush-type polyelectrolyte, it has similarities. Despite the previous comment, and because we do not know a priori the number of binding sites on the copolymer chain, we will calculate the degree of binding as usual, β = $c_b/c_{eq}$, being $c_b$ the concentration of bound surfactant molecules and $c_{eq}$ the concentration of charged monomers in the polyelectrolyte chains. The concentration of charged monomers in PECop solutions at $c_p$ = 400 mg. $L^{-1}$ is 1.65 mM and this value was used in Figure 9.

In Figure 8, for alginate/DTAB, β was calculated using the concentration of charged monomers for the alginate, which is $c_{eq}$ = 2.3 mM. The β values in Figures 8 and 9 will then represent the number of surfactant molecules attached to the polymer chains per charged monomer.

It is evident from Figures 8 and 9 that the number of bound surfactant molecules are higher than the 1:1 charge neutralization ($c_{eq}$) for both systems. For example, for PECop/DTAB mixtures at T2, there are approximately three DTAB molecules per charged group associated with the polymer chain. However, note that the ζ-potential becomes zero at a total surfactant concentration of about 15 mM (Figure 7b, T = 25 °C), which corresponds to the surfactant *cmc* and to $c_b$ = 7 mM. From this, it seems, as we said before, that the association process is not only driven by electrostatic interactions. As we said previously, because the charge inversion occurs at concentrations of bound surfactants five or six times higher than the number of charged groups on the copolymer (see β values on Figure 9), some of the DTAB counterions (Br⁻) must be condensed on (into) the polymer/surfactant aggregates. Note that the charge inversion occurs only at a total surfactant concentrations close and above the *cmc* of the surfactant (~15 mM, Figure 7). All this indicates that the PECop/DTAB association at $c_s$<*cmc* is partially driven by hydrophobic interactions. The previous analysis is also valid for alginate/DTAB complexes (see Figure SI-4). This picture is quite different from what has been found for other homopolyelectrolyte/surfactant [5,60] and polyelectrolyte-copolymer/surfactant mixtures.[21,26,61] in which the structure of the aggregates is compatible with surfactant micelles decorated with polymer chains, bound together via electrostatic interactions between the charged micelles and the oppositely charged groups on the polyelectrolyte chains.

The inversion of the sign in the response when temperature goes over the LCST at high surfactant concentrations, observed by DLS in Figure 4, suggests a change in the association process, and in the structure of the aggregates. At this respect is worth noting that, close to T2 concentration, an inflection point seems to be present in the binding isotherm of Figure 9, suggesting a change in the association regime. Below T2 the slope of the binding isotherm indicates a non-cooperative process (Region 1) while at concentrations above T2, the amount of surfactant molecules bound to the polymer chain increases, showing a more cooperative (not much) binding process (Region 2). This behaviour would indicate, as stated above, a change in the association process and the structure of aggregates, probably due to the presence of a few micelles decorating the aggregates. However, for $c_s$ > *cmc*, the systems are outside the limits of applicability of eqs. 4 and 5, and the data should be taken with caution for concentrations above the *cmc*. However, it is worth noting that the size increase observed when T passes above LCST for PECop/DTAB systems is similar to what happens in alginate/DTAB mixtures as the concentration of surfactant increases above the *cmc*.

The simplest binding isotherm model to analyse the data on Figures 8 and 9 is the Satake−Yang equation[62], that we use here slightly modified[49],

$$\beta = \frac{A}{2}\left[1 + \frac{yc_f - 1}{\sqrt{\left((1-yc_f)^2 + \frac{4yc_f}{u}\right)}}\right] \quad (6)$$

Where y = K$\upsilon$ is the binding constant between a surfactant ion and a site adjacent to a site already occupied by a surfactant, $\upsilon$ is a cooperativity parameter which is determined by the hydrophobic interaction between two adjacently bound surfactants, K is the binding constant between the surfactant and an isolated polyion binding site, and A is the maximum bound fraction. The concentration of free surfactant, and the slope of the binding curve, at the half-bound point, β=A/2, are given by,

$$c_f(\beta = A/2) = \frac{1}{Ku}; \quad \frac{d\beta}{d\ln(c_f)}(\beta = A/2) = \frac{\sqrt{v}}{4} \tag{7}$$

The lines shown in Figures 8 and 9 are the results of equation (6) using the parameters summarized in Table 1, for both alginate/DTAB and PECop/DTAB mixtures. Note that we divided the PECop/DTAB binding isotherm in two regions and analysed them as they were independent. We also fitted the whole isotherm with equation (6) ignoring the inflection point. Equation (7) permits us to estimate the critical aggregation concentration, *cac* by taking the concentration of free surfactant at the half-bound point, $cac_f \sim 1/Kv$. For PECop/DTAB mixtures we have $cac_f$= 0.91 mM for Region 1, which corresponds to $c_s \sim 4$ mM (see Figure 9). This concentration falls in the middle of the first plateau in the surface tension isotherm (Figure 1), between T1 and T2. For Region 2 we obtained $cac_f$ = 4.45 mM which corresponds to $c_s \sim 12$ mM (see Figure 9), this concentration is on the second plateau observed in the surface tension isotherm, and very close to T3.

*Table 1: Parameters of the Satake-Yang empirical equation (eq.(6))*

|  | A | $Ku$ (mM$^{-1}$) | $v$ | $1/Kv$ | $\frac{\sqrt{v}}{4}$ |
|---|---|---|---|---|---|
| Alginate/DTAB | 4.9 ± 0.1 | 0.144 ± 0.005 | 2.9 ± 0.7 | 6.9 | 0.43 |
| PECop/DTAB Region 1 | 3.3 ± 0.3 | 1.1 ± 0.1 | 3.2 ± 0.5 | 0.91 | 0.45 |
| PECop/DTAB Region 2 | 8.1 ± 0.2 | 0.23 ± 0.01 | 0.33 ± 0.04 | 4.45 | 0.14 |
| PECop/DTAB | ~12.7 | 0.096 ± 0.01 | 0.50 ± 0.06 | 10.4 | 0.18 |

Now, comparing the values obtained for A, the maximum bound fraction, for alginate/DTAB and for Region 1 in the PECop/DTAB binding isotherm, we have 4.9 and 3.3 respectively, indicating that the DTAB molecules bound to PECop are about 70% of the surfactant bound to the alginate. This roughly matches the fraction of substitution of the PNIPAAm chains in the copolymer (see Material section). If this is correct, it seems plausible that Region 1 corresponds mainly to the surfactant binding to the alginate chain, close to the charged groups, a conclusion previously stated. Since the maximum value of A for PECop is approximately 12, and for alginate close to 5, the fractional increase in load capacities is ~240%. However, this occurs in the range of surfactant concentration where for which eq. 4 and 5 are out of the range of validity. For $c_s \sim 1$ mM and bellow, when the systems can still respond thermally, the loading is given by the β values at that concentration, from Figures 8 and 9, we have β~0.25 for alginate and β~1.5 for PECop, being the increase in load capacities of 600%.

## 5. CONCLUSIONS

We studied a graft "co-polyelectrolyte" with a brush-type structure mixed with an oppositely charged surfactant as a function of surfactant concentration and temperature. By means of dynamic and static light scattering, we found, for $c_s < cmc$, that the addition of DTAB for T < LCST produces a continuous collapse of the polymer chain, similar to what happens when the temperature is increased to values above LCST, in the absence of the surfactant. This behavior is opposite to that of PNIPAam/SDS.[56] At concentrations above the DTAB *cmc*, the aggregates increase in size rather than decrease in size when temperature crosses LCST. We found that the aggregates formed are quite monodisperse, although the polymer size distribution is broad, a fact that has been rarely observed before.[57]

From mobility and ζ-potential measurements, we constructed the binding isotherms and found that presents an inflection point dividing the isotherm in two regions. The inflection point occurs at a concentration close to T2, indicating two successive binding processes or mechanisms, the first one seems to correspond to the association of DTAB molecules onto the alginate chain, driven partially by electrostatic interactions, the second process is mainly driven by hydrophobic interactions. As a result, the binding isotherm deviates from the common sigmoidal shape; it has the shape of two consecutive and superimposed sigmoidal curves. In this sense, the curve is a mixture of those typical for oppositely charged polyelectrolyte/surfactant mixtures, like that of alginate/DTAB, and those for surfactants binding to hydrophobically modified polymers.[1,59,63] The whole binding process for PECop/DTAB complexes is gradual, being the cooperativity constants very small. Because the sign of the ζ-potential changes at very high surfactant concentration (~15 mM for PECop and ~13 mM for alginate), we conclude that a fraction of the DTAB molecules bound to the polymer chain do so with their counterions.

The aggregates respond to temperature changes only within certain surfactant concentrations. For $c_s$ < 1 mM, the aggregates decrease their size when temperature increases above LCST. On the contrary, for $c_s$ > 10 mM, the aggregates increase their size when T > LCST. No appreciable response is observed at concentrations in the range 1 < $c_s$ < 10 mM, being the minimum response at $c_s$ = 2.8 mM. This change of ¨sign¨ in response to temperature, as well as the loss of responsiveness depending on surfactant concentration, is quite unique. We have shown that the amount of DTAB bound to PECop is 600% that of alginate, over the range of surfactant concentration where the copolymer can still respond to changes in temperature.

The behavior shown by the PECop/DTAB mixtures presented here could be relevant for the design of drug delivery systems based on mixtures of this co-polymer with other small molecules.

**Acknowledgements**

This research was funded by the Agencia Nacional de Promoción Científica y Tecnológica (AN-PCyT), Argentina, PICT-2019-3185, by Consejo Nacional de Investigaciones Científicas y Técnicas (CONICET), Argentina, under grant PIP 11220200101754CO, and by the Universidad Nacional del Sur (UNS), Argentina, PGI-UNS 24/F080.